\documentclass[aps,prl,amsmath,amssymb,reprint,groupedaddress]{revtex4-1}
\usepackage{graphics}
\usepackage{epsfig}
\usepackage{color}
\usepackage{epstopdf} %epd ͼƬ
\usepackage{bm}  %����
    \newcommand{\ba}{\begin{eqnarray}}
    \newcommand{\ea}{\end{eqnarray}}
    \newcommand{\be}{\begin{equation}}
    \newcommand{\ee}{\end{equation}}
    \newcommand{\calM}{{\mathcal M}}
\begin{document}

%\begin{CJK*}{GB}{} % Use default fonts from CJK (see below)
\title{Three Photon Decay of $J/\psi$ from Lattice QCD}

\author{Yu Meng$^{1}$, Chuan Liu$^{1,2}$, Ke-Long Zhang$^{1}$}

\affiliation{
$^1\,${\it School of Physics and Center for High Energy Physics,}\\
{\it Peking University, Beijing 100871, P.R. China}\\
$^2\,${\it Collaborative Innovation Center of Quantum Matter,}\\
{\it Beijing 100871, P.R. China}\\
}

\date{\today}

\begin{abstract}
%  We propose a method to directly study the distribution of decay rate in the Dalitz plot for three-body decays
%  using lattice QCD methods. This opens up the possibility of studies on three-body decays involving strong interactions using nonperturbative theoretical methods.
  Three photon decay rate of $J/\psi$ is studied using two
  $N_f=2$ twisted mass gauge ensembles with lattice spacings $a\simeq 0.085$ fm (I) and $0.067$ fm(II).
  Using a new method, only the correlation functions directly related to the physical
  decay width are computed with all polarizations of the initial and final states summed over.
  Our results for such rare decay on the two ensembles are: $\mathcal{B}_{I,II}(J/\psi\rightarrow 3\gamma)=(1.614 \pm 0.016 \pm 0.261)\times 10^{-5},(1.809 \pm 0.051 \pm 0.295)\times 10^{-5}$ where
  the first errors are statistical and the second are estimates from systematics.
  We also propose a method to analyze the Dalitz plot of the corresponding process
  based on the lattice data which can provide direct information for the experiments.
%  In principle, our new method is applicable to other hadronic decays with more particles in the final states.
%  \vspace{5cm}
\end{abstract}

%\keywords{}

\maketitle
%\end{CJK*}

\emph{Introduction} -- The rare decay $J/\psi \rightarrow 3\gamma$, analog to Ortho-positronium decaying to $3\gamma$ in quantum electrodynamics (QED)~[\onlinecite{QED_test}], can provide a high precision test for the non-perturbative quantum chromodynamics (QCD)~[\onlinecite{QCD_nonper}] because of the charmonia scale in strong interaction~[\onlinecite{QCD_Jpsi}]. Despite decades of effort, such a rare decay had not been observed by experimentalists until 2008, the CLEO collaboration measured the branching fraction $\mathcal{B}(J/\psi\rightarrow 3\gamma)=(1.2\pm 0.3 \pm 0.2)\times 10^{-5}$ for the first time~[\onlinecite{CLEO2008}]. With the help of much larger $J/\psi$ samples, BESIII Collaboration obtained a more accurate result $(11.3\pm 1.8 \pm 2.0)\times 10^{-6}$ in 2013~[\onlinecite{BESIII2013}]. Both the unavoidable system errors come from the uncertainty of the number of $\psi(3686)$ and decay process $\mathcal{B}(\psi(3686)\rightarrow \pi^{+}\pi^{-}J/\psi)$, which undertook the most data set contributions.

On the theoretical side, using perturbative methods,
the modern tools of treating the quarkonium physics is nonrelativistic QCD (NRQCD) factorization~[\onlinecite{NRQCD1995}],
in which the decay rate of $J/\psi \rightarrow 3\gamma$ is parameterized in terms of lowest order NRQCD $J/\psi$-to-vacuum matrix element plus relativistic corrections $\langle v^2\rangle_{J/\psi}$~[\onlinecite{jiayu2013}]. However, when going to higher orders,  both the inconsistency between theory and experiments and divergence puzzle in higher order radiative corrections~[\onlinecite{divergent_nrqcd}] indicate that the NRQCD may break down for predicting $J/\psi \rightarrow 3\gamma$ decay rate. Therefore, it is fair to say that,
even after three decades, the understanding of the process $J/\psi \rightarrow 3\gamma$ within NRQCD has not improved much when compared with the situation in early 1980s~[\onlinecite{potential_1,potential_2,potential_3}].

So, it is of great significance to seek new methods. In this letter, we propose to use lattice QCD as such an alternative and we present the first exploratory computation of $J/\psi \rightarrow 3\gamma$ decay width using
 two ensembles of gauge field configurations. Using lattice QCD, one usually evaluates the matrix element of interested interpolating operators with correct quantum numbers between hadronic states. Although photon itself is not an eigenstate of QCD, regarding the photon as a superposition of QCD eigenstate and adoping the electromagnetic current $J^{\mu}_{em}$ as photon interpolating operators has been proposed already ~[\onlinecite{Ji2001}] which has been widely used in photon structure functions~[\onlinecite{Ji2001_2}], radiative transition ~[\onlinecite{Dudek2006_2}] and two-photon decays in charmonia ~[\onlinecite{Dudek2006},\onlinecite{CLQCD_etac}].

\emph{Method} -- We start by expressing the amplitude of $J/\psi \rightarrow 3\gamma$ in terms of the appropriate four-point function using Lehmann-Symanzik-Zimmermann reduction formula, integrating out the photon fields perturbatively and continuing the resulting expression to Euclidean space analytically. This process introduces the photon virtualities $Q_i^2=|\bm{q}_i|^2-\omega_i^2$ (for more details, see Ref.~[\onlinecite{Dudek2006}]),
we then arrive at the following final result for the four-point function
that is relevant for the process $J/\psi\rightarrow 3\gamma$,
\begin{widetext}
\begin{eqnarray}\label{main-result1}
  %%&& M(t_f,t;t',t_i)\equiv \frac{1}{V\cdot T}\langle J/\psi(p,\lambda_0) \vert \gamma(q_1,\lambda_1)\gamma(q_2,\lambda_2)\gamma(q_3,\lambda_3)\rangle \nonumber \\
  && M(t_f,t;t',t_i)=  \lim\limits_{t_f-t\rightarrow \infty} e^{3}
  \frac{
  \epsilon_{\mu}(q_1,\lambda_1)\epsilon_{\nu}(q_2,\lambda_2)\epsilon_{\rho}(q_3,\lambda_3)\epsilon_{\alpha}(p,\lambda_0)
  }
  {\frac{Z_{J/\psi}(\bm{p})}{2E_{J/\psi}(\bm{p})}e^{-E_{J/\psi}(\bm{p})(t_f-t)}}
  \int dt^{'} e^{-\omega_2|t^{'}-t|} \int dt_i e^{-\omega_1|t_i-t|} \nonumber \\
  &&\times \left\langle 0 \left \vert T\left\{\mathcal{O}^\alpha_{J/\psi}(\bm{0},t_f) \int d^{3}\bm{z}e^{i\bm{q}_3\cdot\bm{z}}j^{\rho}(\bm{z},t)\int d^3\bm{y}e^{i\bm{q}_2\cdot\bm{y}}j^{\nu}(\bm{y},t^{'})\int d^3\bm{x}e^{i\bm{q}_1\cdot \bm{x}}j^{\mu}(\bm{x},t_i)\right\} \right \vert 0\right\rangle\;.
\end{eqnarray}
\end{widetext}
Here the four polarization vectors: $\epsilon_\mu$, $\epsilon_\nu$, $\epsilon_\rho$ and $\epsilon_\alpha$
correspond to the three final photons and the initial $J/\psi$ particle, respectively, with
the polarizations labelled by $\lambda_1$, $\lambda_2$, $\lambda_3$ and $\lambda_0$.
The analytic continuation from Minkowski to Euclidean space here works out as long as the virtualities of three photons are not too time-like to produce on-shell vector hadrons. More specifically, $Q_i^2 = \vert \bm{q}_i\vert^2-\omega_i^2 > -M_{V}^2$ where $M_{V}$ is mass of the lightest vector meson.
The correlation functions appearing in the above equation can be evaluated in lattice QCD
in terms of quark propagators. In this exploratory calculation, we have neglected the disconnected diagrams.
\begin{figure}[htb]
\includegraphics{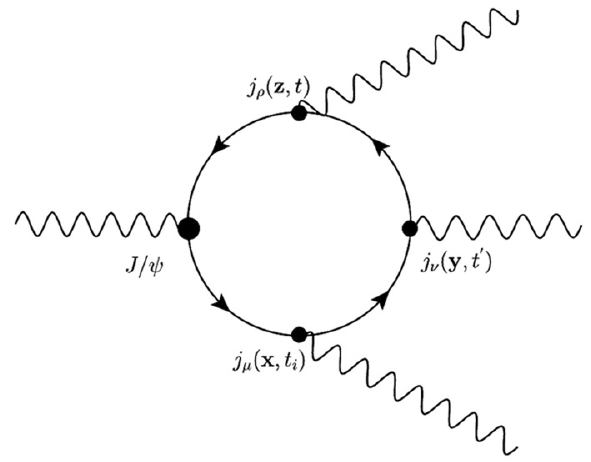}% Here is how to import EPS art
\caption{\label{Jsi_3photon} Connected diagram computed for the process $J/\psi \rightarrow 3\gamma$.}
\end{figure}
For simplicity, we denote the matrix element in Eq.~(\ref{main-result1}) as $M=\epsilon_{\mu}\epsilon_{\nu}\epsilon_{\rho}\epsilon_{\alpha}\mathcal{M}_{\mu\nu\rho\alpha}$ and introduce $T \equiv \overline{|M|}^2=\frac{1}{3}\sum_{\mu\nu\rho\sigma}|\mathcal{M}_{\mu\nu\rho\sigma}|^2$ which will be called $T$-function in the following.
As we will see, $T$ function represents a distribution
of physical partial decay width in terms of a pair of kinematic variables.
Each $\mathcal{M}_{\mu\nu\rho\alpha}$ can be computed on the lattice using the fact that
 $M$ is independent of the time $t$, as long as $|t_f-t|$ is large enough.
 For simplicity, we have used the local current $j_\mu(x)=\bar{c}(x)\gamma_\mu c(x)$
 for the charm quark which can be renormalized by a multiplicative factor $Z_V$.
 In real simulations, plateau behaviours are searched for to extract the values of $M$ in various cases.
 The current coupling to first photon is fixed at $t_{i}$, other two are placed at $t^{'}$ and $t$, respectively and $J/\psi$ meson is fixed at $t_f$(as shows in Figure.~\ref{Jsi_3photon}). The integrals in Eq.~(\ref{main-result1}) are also replaced by corresponding trapezoidal summations.
%To really evaluate the matrix element $M$, we form an appropriate ratio of
%the four point functions with respect to the two point function,
%\be
%\tilde{M}=\frac{M(t_f,t;t',t_i)}{\frac{Z_{J/\psi}(\bm{p})}{2E_{J/\psi}(\bm{p})}e^{-E_{J/\psi}(\bm{p})(t_f-t)}}
%\;.
%\ee
% When the time-separation is large, the above ratio saturates to the constant
% of the transition matrix element for $J/\psi\rightarrow 3\gamma$.

 In conventional lattice computations, for example in the decay of $\eta_c\rightarrow\gamma\gamma$ etc.,
 the hadronic matrix element such as $\mathcal{M}_{\mu\nu\rho\alpha}$ is
 further decomposed into various form factors which are functions of the virtualities $Q^2_i$.
 By fitting the matrix element at different $Q^2_i$ with a particular functional form,
 one arrives at the complete off-shell form factors and finally
 the physical decay width can be obtained by setting all virtualities to the on-shell values, namely $Q^2_i=0$,
 yielding the final decay rate.
 In our case, the form factor decomposition is way too complicated. In the study of
 $3\gamma$ decays of $Z$ and the positronium, people have worked out the decomposition
 in perturbation theory~[\onlinecite{Z_boson},\onlinecite{ee_3photon}]. However, there is no guarantee that
 these perturbative decomposition will also work in QCD. Therefore, we will proceed in another way.
 We will be satisfied with the physical decay width only. That is to say, we will be only interested in the
 on-shell matrix element. Thus, we can perform the summation over polarizations of the initial and final
 particles first, and only the on-shell matrix element will be computed on the lattice.
 Due to Ward-identities of the currents, the summation over polarizations of the photons
 yields the Minkowski metric, e.g. $\sum_{\lambda_i}(\epsilon_\mu(q_i,\lambda_i)\epsilon^*_{\mu'}(q_i,\lambda_i)\Rightarrow -g_{\mu\mu'}$.
 The summation over the initial polarization of $J/\psi$ yields the same if we take the rest frame
 of the particle. Therefore, we have $\sum_{\lambda_i} |\mathcal{M}|^2=\sum_{\mu\nu\rho\alpha} |\calM_{\mu\nu\rho\alpha}|^2$.
 In our actual simulations, we sum over all polarizations (altogether 192 possibilities) of $\mathcal{M}$.
% Therefore, we only need to evaluate the quantity $T$ in the following.

 The decay width of $J/\psi \rightarrow 3\gamma$ in $J/\psi$ center of mass frame can be expressed as,
\begin{widetext}
\begin{eqnarray}\label{decay-width}
\Gamma(J/\psi\rightarrow3\gamma)=&&\frac{1}{3!}\frac{1}{2M_{J/\psi}}\int\frac{d^3q_1}{(2\pi)^32\omega_1}\frac{d^3q_2}{(2\pi)^32\omega_2}\frac{d^3q_3}{(2\pi)^32\omega_3}(2\pi)^4\delta(p-q_1-q_2-q_3)\overline{|\mathcal{M}|^2} \nonumber\\
=&& \frac{m_{J/\psi}}{1536\pi^3}\int_{0}^{1}dx \int_{1-x}^{1}dyT(x,y)
\end{eqnarray}
\end{widetext}
where $x,y$ are two dimensionless variables in the range $[0,1]$,
defined as $x\equiv 1-2q_2\cdot q_3/M^2_{J/\psi},y\equiv 1-2q_1\cdot q_2/M^2_{J/\psi}$.
 It is easily checked that they fall into the right-upper triangle of
 the unit square in the $xy$-plane, i.e. satisfying : $x\in [0,1], y\in [1-x,1]$.
% Remind that the T-amplitude in Eq.(~\ref{decay-width}) is on-shell.
 In the continuum, the on-shell decay pattern are normally parameterized by the so-called Dalitz plots,
 which can be obtained from the $T$-function $T(x,y)$.
 Due to the discreteness of the momenta on the finite lattice, it is
 impossible to exactly impose on-shell condition for all particles, making the on-shell quantity $T(x,y)$
 not directly accessible. Instead, the on-shell conditions for the particles can be done
 as follows: We first put the $J/\psi$ particle and at least one final photon on shell, keeping
 the other two photons as close to on-shell as possible by adjusting their three momenta.
 It is found that this still introduce some non-vanishing virtualities to the other photons.
 With these non-vanishing but small virtualities, the matrix element can be computed directly on the lattice,
 the norm of which we denote as $T(x,y,Q_1^2,Q_2^2,Q_3^2)$. This differs from the $T$-function only
 because of the fact that some of the photons are still not on-shell.
 We then try to estimate the on-shell quantity, the $T$-function $T(x,y)$,
 by the following fitting formula,
\begin{equation}\label{on-shell_formula}
T(x,y,Q_1^2,Q_2^2,Q_3^2)= T(x,y)+ \text{const} \times\sum\limits_i Q_i^2
\end{equation}
for $\vert Q_i^2\vert \ll 1$ where everything is measured in lattice units.
We expect such behavior since the final three photons are identical.

The Eqs.~(\ref{main-result1}),\ (\ref{decay-width}) and (\ref{on-shell_formula}) constitute the central part of this letter. As pointed out already, different from the conventional method, we have intentionally avoided the amplitude parameterization for $J/\psi \rightarrow 3\gamma$, though it has the similar structure as $Z\rightarrow 3\gamma$~[\onlinecite{Z_boson}] and the positronium to $3\gamma$ decay~\cite{ee_3photon}. Because all the form factors introduced in the amplitude parameterizations are scalar functions of three-photon momenta, permutations of these momenta then lead to more form factors, rendering the computation of all of these form factors too costly.
In the case of three-body decay, what is really measured in the experiments are the so-called
Dalitz plots of the final states. Dalitz plot represents the distribution of the partial
decay width in two independent kinematic variables. In the case of three-photon decay
of $J/\psi$, this is taken to be the largest and the smallest two-photon invariant mass
values, denoted as $M(\gamma\gamma)_{lg}$ and $M(\gamma\gamma)_{sm}$, respectively,
among three combinations for the final photons.
 We will call them the Dalitz variables in the following.
These two Dalitz variables are directly related to the kinematic
variables $(x,y)$ that we introduced. To be more specific, we have,
\be
\label{eq:relation_M_and_xy}
\!\!\!\frac{M(\gamma\gamma)_{lg/sm}}{M_{J/\psi}}=\begin{aligned}
\max\\
\min\end{aligned}\left\{\sqrt{1-x},\sqrt{1-y},\sqrt{x+y-1}\right\},
%M(\gamma\gamma)_{sm}=&M_{J/\psi}\times \min\left\{\sqrt{1-x},\sqrt{1-y},\sqrt{x+y-1}\right\}
\ee
%\textcolor{red}{Maybe we give out the relation directly?}
where the upper/lower line on the right corresponds
to the case of $M(\gamma\gamma)_{lg}$/$M(\gamma\gamma)_{sm}$, respectively.
Thus, the Dalitz plot for the three-body decay is directly related
to the on-shell $T$-function $T(x,y)$ that we aim to compute on the lattice.

\emph{Simulations And Results} -- Our lattice calculation is performed using two $N_f=2$ flavour twisted mass gauge field ensembles generated by the Extended Twisted Mass Collaboration (ETMC) with lattice spacing $a \simeq 0.067$ fm and $0.085$ fm, respectively~[\onlinecite{ETMC2007}]. Relevant information for these are listed in Table.~\ref{table:cfgs}.

\begin{table}[!h]
\caption{\label{table:cfgs}%
 Information for the gauge ensembles.}
\begin{ruledtabular}
\begin{tabular}{ccccccc}
\textrm{Ensemble} & $\beta$  & $a$(fm) & $V/a^4$  & $a\mu_{\textrm{sea}}$ & $m_{\pi}$(MeV) & $N_{\textrm{conf}}$\\
%\multicolumn{1}{c}{\textrm{Three}}& \\
\hline
I     &  3.9       &  0.085   & $24^3\times 48$ & 0.004 & 315  &40 \\
II    &  4.05      &  0.067   & $32^3\times 64$ & 0.003 & 300  &20 \\
\end{tabular}
\end{ruledtabular}
\end{table}

\begin{figure}[ht]
\includegraphics{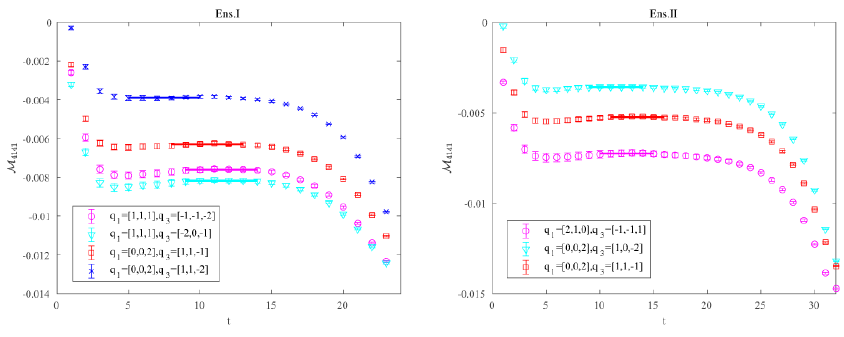}% Here is how to import EPS art
\caption{\label{M4141} Four-point function $\mathcal{M}_{\mu\nu\rho\alpha}$ as function of $t$. 4 sets of photon momenta in Ens.I(left) and 3 sets in Ens.II(right) are chosen to ensure discrete data points of $T(x,y)$ can
 cover the integral region $x \in (0,1), y \in (1-x,1)$  under
 the condition that all virtualities are small.\label{fig:plateau}}
\end{figure}

The conventional sequential method has been adopted to calculate the four-point functions. Two sequential sources are placed close to $J/\psi$ meson, and the contraction is performed on the furthest current. After the integration (summation) of time slice $t_i$ and $t^{'}$, the matrix element $\mathcal{M}_{\mu\nu\rho\alpha}$, being a function of time slice t, can be obtained on the lattice. We have chosen 4 sets of photon three-momenta with suitable photon virtualities $Q_i^2$ in Ens.I and 3 in Ens.II . In Fig.~\ref{M4141}, typical plateau behaviors for the four-point function $\mathcal{M}_{\mu\nu\rho\alpha}$ are shown in the case of $\mu\nu\rho\alpha=4141$.
The data points with errors are the results from the simulation and the errors are estimated using
jackknife method. Other cases are similar.

\begin{figure}[!h]
\includegraphics{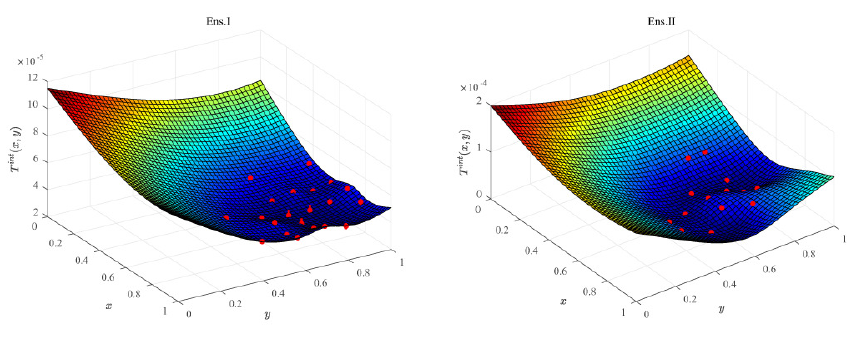}% Here is how to import EPS art
\caption{\label{T_surface} The interpolated $T$-function $T^{(int)}(x,y)$ are shown.}
\end{figure}

As is seen from Fig.~\ref{fig:plateau}, although only 4 sets of photon three-momenta in Ens.I and 3 in Ens.II are considered,
the physical amplitude %$\langle J/\psi(p,\lambda_0) \vert \gamma(q_1,\lambda_1)\gamma(q_2,\lambda_2)\gamma(q_3,\lambda_3)\rangle $
is invariant under the photon exchange $(\bm{q}_i,\lambda_i) \leftrightarrow (\bm{q}_j,\lambda_j)$, so we finally obtain the off-shell
$T$-function, i.e. $T(x,y,Q_1^2,Q_2^2,Q_3^2)$, at a total of 21 and 15 points of $(x,y)$ in the $xy$-plane, respectively.
For each of these, the on-shell function $T(x,y)$ can be extracted by performing a correlated fit
using Eq.~(\ref{on-shell_formula}) with bootstrap method. Finally, we utilize a cubic spline function to interpolate these on-shell $T(x,y)$ points. The surface of these resulting interpolating functions $T^{(int)}(x,y)$ are illustrated in the Fig.~\ref{T_surface}, together with the original data points shown in red.
%Before we integrate the T-amplitude $T(x,y)$ in ($x-y$) plane to obtain decay width given by Eq.~(\ref{main-result1}), we pause for a moment with the following comments.

 For three-body decays, the decay width is a quantity that
 both experimentalists and theorists are interested in.
 Most of the time, however, the total decay width itself is not directly measurable in experiments.
 Instead, the Dalitz plot, which is a distribution of the decay width
 in the plane of two kinematic variables, is obtained first.
 In the case of $J/\psi\rightarrow 3\gamma$,
 these are exactly the Dalitz variables  $M(\gamma\gamma)_{lg}$ and $M(\gamma\gamma)_{sm}$
 that we mentioned, which are related to the $(x,y)$ kinematic variables via Eq.~(\ref{eq:relation_M_and_xy}).
 Being first-hand data obtained in experiments,
 Dalitz plot plays a key role for a three-body final state.
 As is well-known, bands that appear in the Dalitz plot indicate that there is an intermediate two-body state. Thus, nonuniformity in the Dalitz plot can offer immediate information on the cross section $|\mathcal{M}|^2$.
 As we will illustrate below, this can be related to the on-shell $T$-function that we compute on the lattice.

 %For this purpose, we project the $T(x,y)$ onto $x-y$ plane and it can be converted into Dalitz plot directly. FIG.~\ref{T_surface} is the realization of this idea and also a central result of this letter.
\begin{figure}[!h]
\vspace{3mm}
\includegraphics{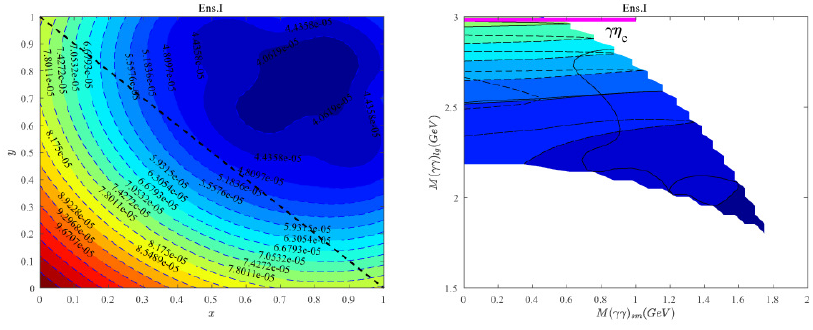}% Here is how to import EPS art
\caption{\label{T_dalitz} The contour plot of Ens.I is shown
  in both $(x,y)$ variables (left) and in the corresponding Dalitz variables (right), the two are related by Eq.~(\ref{eq:relation_M_and_xy}). Plots for Ens.II are similar.}
\end{figure}

 It is seen that, the relation between the Dalitz variables and the pair $(x,y)$ as
 indicated in Eq.~(\ref{eq:relation_M_and_xy}), maps the upper right triangular region of unit square
 in the $(x,y)$ plane onto a corresponding region in $(M(\gamma\gamma)_{sm},M(\gamma\gamma)_{lg})$ plane
 in the Dalitz plot. The shape of the region in the Dalitz variables is not regular
 but this is exactly what is measured in the experiments, see e.g. Fig.1 (d) in Ref.~\cite{BESIII2013}.

% We illustrate this in the bottom half of Fig.~\ref{T_surface}.
 As we have obtained the interpolating functions $T^{(int)}(x,y)$ illustrated in the Fig.~\ref{T_surface}, we illustrate the mapping from $(x,y)$ plane to the
 Dalitz variables plot as suggested in Eq.~(\ref{eq:relation_M_and_xy}).
 This is shown in Fig.~\ref{T_dalitz} in the case of Ens.I.
 On the left is the contour plot of the interpolated function $T^{(int)}(x,y)$ while on the right is
 the corresponding one in Dalitz variables.
 To further obtain the total decay width for the process, one needs to either integrate
 the function $T(x,y)$ in the $(x,y)$ plane, or doing the corresponding integration in the Dalitz variables.

% \textcolor{blue}{
% Here, I pause for a while, until we clarify the issue whether our computation also
% includes $J/\psi\rightarrow\gamma\eta_c\rightarrow 3\gamma$. If this were the case,
% we should follow what the experimentalists are doing, cutting out the processes
% that belong to the final states $\gamma\eta_c$.
% }

 Before we integrate the function to give the final decay rate, let us make the following comments:
\begin{itemize}

\item[i)] We have computed the connect diagram as shown in Fig.~\ref{Jsi_3photon}. This diagram can
 in principle also include the physical process $J/\psi\rightarrow\gamma\eta_c\rightarrow\gamma\gamma\gamma$ as well. Therefore, in order to make comparison with the experiments, we need to remove such contributions
 from our lattice data. It is easily verified that this corresponds to the corners of the triangle in
 the $(x,y)$ plane. In the experiments, these are also the regions where the major background comes in.
 To remove these contributions, we need to make definite cuts as the experimentalists did, see e.g. Ref.~[\onlinecite{CLEO2008},\onlinecite{BESIII2013}]. For example, we cut the three corners by the
 condition $M(\gamma\gamma)_{lg}<2.9$ GeV, resulting in a deduction of $0.031$ eV in Ens.I and $0.034$ eV in Ens.II in the final
 results for $\Gamma(J/\psi\rightarrow 3\gamma)$ shown in Eq.~(\ref{width}) below, which are from the region $x \in [0,0.1],y \in [1-x,1]$ and $x \in [0.9,1],y \in [1-x,0.1]$ both in two ensembles.

\item[ii)] No obvious bands found on the vertical region, especially for the range $M(\gamma\gamma)_{sm}(GeV)$ $\in$  $[0.1,0.16]$,$[0.5,0.6]$,$[0.9,1](GeV)$, which correspond to the dominant sources $\gamma\pi_0,\gamma\eta, \gamma\eta^{'}$ in experiments~[\onlinecite{CLEO2008},\onlinecite{BESIII2013}]. This is understandable because such contributions are excluded in the connected diagram of $J/\psi$ decay in Fig.~\ref{Jsi_3photon}.
\end{itemize}

 Now we proceed to integrate $T$-function surface over the physical region
 and finally arrive at the decay width of $J/\psi \rightarrow 3\gamma$,
 \begin{align}\label{width}
\Gamma_{I}(J/\psi\rightarrow 3\gamma) &= 1.499(15)(243)\ \mbox{eV}\; \nonumber \\
\Gamma_{II}(J/\psi\rightarrow 3\gamma)&= 1.681(47)(274)\ \mbox{eV}\;.
\end{align}
 Here the first errors are statistical and second are our estimates for the systematics.
 The statistical ones contain the errors from current renormalization factor $Z_V^{I,II}=0.6347(26),0.6640(27)$
 which are computed using a ratio of three-point function over two-point functions
 as in Ref.~\cite{Dudek2006_2} and the errors from on-shell fitting process as suggested in Eq.~(\ref{on-shell_formula}).
 The systematic errors are  from the cubic spline interpolation process, which are obtained by estimating the integrating results of $T^{(int)}(x,y)$ in $xy$-plane without original data points, in particular in the region of $x \in [0.1,0.3],y \in [1-x,1]$ and $x \in [1-y,1],y \in [0.1,0.3]$ for both ensembles.

 The branching fraction, if the uncertainty of $J/\psi$ total width being ignored, is given by $\mathcal{B}_{I,II}(J/\psi\rightarrow 3\gamma)=(1.614 \pm 0.016 \pm 0.261)\times 10^{-5},(1.809 \pm 0.051 \pm 0.295)\times 10^{-5}$,
 which are consistent with both the results of CLEOc and BESIII Collaboration within $3\sigma$ accuracy.
 We emphasize that, since we have only two lattice spacing values, we cannot make the continuum
 extrapolation in a controlled fashion. This is another source of systematic error that needs
 to be taken into account. However, the two values at two lattice spacings indicate that, this
 is likely within our current estimate for the systematic errors.

\emph{Discussions} -- As we have said, the Dalitz plot is usually the first obtained observable for three-body decays. Therefore, it is instructive to present the Dalitz plot directly which offers a more detailed comparison of the lattice result and the experiment.
 For this purpose, we define a normalized $T$-function distribution density $\tilde{T}(x,y)$ as
\begin{equation}\label{T_density}
\tilde{T}(x,y)=\frac{T(x,y)}{\int_{0}^{1}dx \int_{1-x}^{1}dy T(x,y)}\;,
\end{equation}
which can be viewed as probability density in the $(x,y)$ plane.
The corresponding Dalitz plot can also be generated by drawing random samples using this probability distribution.
\begin{figure}[!h]
\vspace{5mm}
\includegraphics{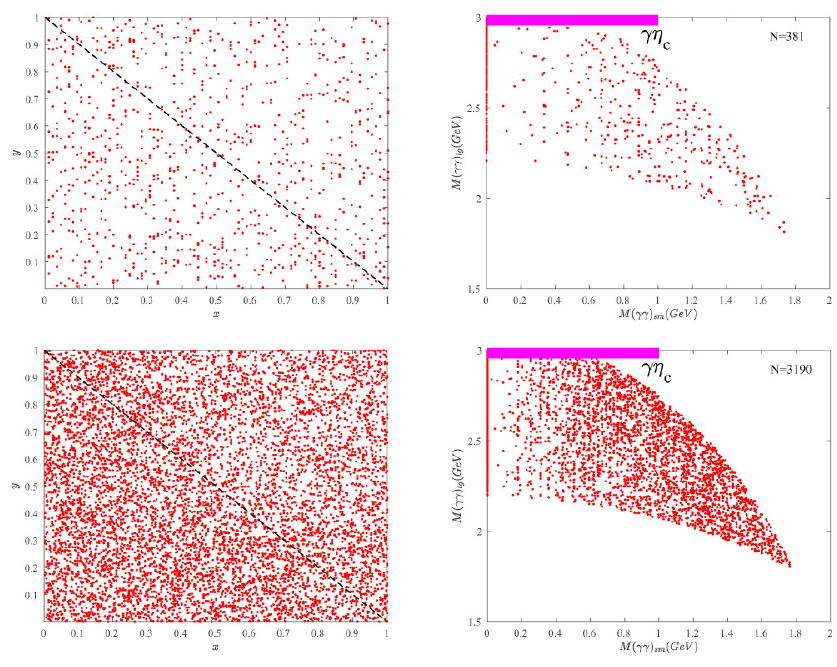}% Here is how to import EPS art
\caption{\label{T_sample} The left panels are the samplings with probability density given in Eq.~(\ref{T_density}), with $N=381$ (top) and $N=3190$ (bottom) samples, respectively.
The right panels are the corresponding Dalitz plots.}
\end{figure}

 Taking results from Ens.I as an example, in Fig.~\ref{T_sample} we illustrate the distribution of data points drawing from
 probability distribution $T(x,y)$ defined in Eq.~(\ref{T_density}) with $N=381$ and $N=3190$ random samples.
 On the left, we show the distribution in $(x,y)$ variables,
 taking $N=381$ (top) and $N=3190$ (bottom) samples, respectively.
 The right panels show the corresponding Dalitz plots.
 Note that the number $N=381$ is almost the same as $J/\psi$ events observed in BESIII.
 The Dalitz plot for this low statistics resembles that in BESIII experiment qualitatively.
 So we expect BESIII would be able to observe the features with higher statistics in Fig.~\ref{T_sample}
 in the future with $1.39\times 10^{9}$ $J/\psi$ events already collected.

 To summarize, the method advocated in this particular lattice calculation can also promote other
 similar lattice computations. By summing over final and initial state polarizations, we obtain directly
 the distribution of the partial decay width in the corresponding Dalitz plot,
 which can be compared  directly with the experiments.
 In principle, we could also keep the information of the initial polarization of the $J/\psi$ particle.
 One could also contemplate  to generalize it to other hadronic decays with three particles in the final state.
 As a side remark, this method can be easily applied to processes like $\eta_c\rightarrow\gamma\gamma$.

\emph{Conslusions} -- Using lattice QCD,
  an exploratory calculation of rare decay rate $\Gamma(J/\psi\rightarrow 3\gamma)$ is presented.
  We obtain the branching fraction $\mathcal{B}_{I,II}(J/\psi\rightarrow 3\gamma)=(1.614 \pm 0.016 \pm 0.261)\times 10^{-5},(1.809 \pm 0.051 \pm 0.295)\times 10^{-5}$ with lattice spacing $a\simeq 0.085$ fm(I) and $0.067$ fm(II), respectively.
  The result is consistent within $3\sigma$ level with the two existing experimental ones from CLEOc and BESIII.

  Instead of parameterizing the matrix relevant matrix element with form factors,
  we evaluate the squared matrix element which is directly related to
  the physical decay width. We could also obtain the distribution of the partial decay
   width in terms of two kinematic variables that is directly related to the Dalitz plot in experiments.
  For the decay $J/\psi\rightarrow 3\gamma$ we also predict the Dalitz plot structure with more statistics,
  which could be tested by BESIII Collaboration in the future.

\begin{acknowledgments}
 The authors would like to thank Prof. Xu Feng at Peking University and
 Prof. Luchang Jin at University of Connecticut for helpful discussions.
 The authors also benefit a lot from the discussions with the members of the CLQCD
 collaboration. The numerical work of were carried out on Tianhe-1A supercomputer
 at Tianjin National Supercomputing Center.
 This work is also supported in part by the DFG and the NSFC through funds
 provided to the Sino-Germen CRC 110 ``Symmetries and the Emergence
 of Structure in QCD'', DFG grant no. TRR~110 and NSFC grant No. 11621131001.
%\vspace{6cm}
\end{acknowledgments}

\clearpage

\end{document}